\documentclass[a4paper,12pt,twoside]{article}
\usepackage{a4wide}
\usepackage[centertags]{amsmath}
\usepackage{amsfonts}
\usepackage{rotating}

\numberwithin{equation}{section}

\let\rho\varrho
\let\phi\varphi

\newcommand{\Spin}{\text{Spin}} 
\newcommand{\pU}{\text{U}  }     
\newcommand{\pE}{\text{E}  }     

\newcommand{\mR}{\mathbb{R}}
\newcommand{\mC}{\mathbb{C}}
\newcommand{\mZ}{\mathbb{Z}}

\newcommand{\CP}{{\mC P}}

\newcommand{\Dd}{\text{d}}      
\newcommand{\Dp}{\partial}      

\newcommand{\auf}{\rightarrow}
\newcommand{\lauf}{\longrightarrow}

\DeclareMathOperator{\tr}{tr}

\DeclareMathOperator{\id}{id}

\newcommand{\qed}{\ \hfill $\Box$\par}

\newcommand\DDS{\text{\hbox to 0pt{/\hss}D}}
\newcommand\DpS{\text{\hbox to 0pt{/\hss}}\partial}

\begin{document}

\begin{titlepage}

\noindent\hfill hep-th/0009251


\vfill

\begin{center}
  \begin{center}
    \Huge On Fractional Instanton Numbers in Six Dimensional Heterotic
    $\pE_8 \times \pE_8$ Orbifolds
  \end{center}

  \vspace{1.5cm}

  Jan O. Conrad\footnote{E-mail: conrad@th.physik.uni-bonn.de}

  \smallskip

  {\it Physikalisches Institut, Universit\"at Bonn\\
    Nu\ss{}allee 12, D-53115 Bonn, Germany}

  \vfill

  {\bf Abstract}
  
  \bigskip

  \parbox{0.9\textwidth}{%
    We derive the precise relation between level matching condition
    and fractional instanton numbers in six dimensional, abelian and
    supersymmetric orbifolds of $\pE_8 \times \pE_8$ heterotic string
    theory. The fractional part of the two $\pE_8$ instanton numbers
    is explicitly calculated in terms of the gauge twist.  This
    relation is then used to show that the classification of these
    orbifolds can be given in terms of flat bundles away from the
    orbifold singularities under the only constraint that the sum of
    the fractional parts of the gauge instanton numbers match the
    fractional part of the gravitational instanton number locally at
    every fixed point. This directly carries over to M-theory on
    $S^1/\mZ_2$.}

\end{center}

\vfill

\end{titlepage}

\section{Introduction and Summary}

Since the $\pE_8 \times \pE_8$ heterotic string theory has been
related to M-theory on $S^1/\mZ_2$ for the first time \cite{HWI,HWII}
(see also \cite{HetTypeI}) some effort has been made to describe
heterotic orbifolds in this setting \cite{Stieberger, OvrutI,
  Kaplunovsky, OvrutII}. However, in the phenomenologically
interesting case of compactification to four dimensions our
understanding is still extremely limited (see \cite{KaplSUSY2K}) and
even in the much simpler case of six noncompact dimensions only some
models have been treated successfully, whereas a large class of models
still lacks a fully satisfying description \cite{Kaplunovsky}.

In the present paper we focus on symmetric, abelian and perturbative
orbifolds of the $\pE_8 \times \pE_8$ heterotic string in six
dimensions preserving 8 supersymmetries as described in \cite{OrbI,
  OrbII} (without discrete torsion).

At first, we consider a single fixed point in an arbitrary model
located at the origin of $\mC^2$. Associated to it is a generator
$(r,\gamma)$ of $\mZ_N$ consisting of a rotation $r$ acting like
$\exp(2\pi i \Phi_i)$ on the complex coordinate $Z^i$ and a gauge
shift $\gamma$ acting like $\exp(2\pi i \beta^I_1 H^1_I) \exp(2\pi i
\beta_2^I H^2_I)$ where the $H_I^{1,2}$ are the eight generators of
the Cartan subalgebra in the adjoint representation of
$\pE_8^{(1,2)}$. The $H_I$ will be normalized such that the $\pE_8$
lattice $\Lambda_8$ is given by the vectors $p^I = (n_1, \ldots, n_8)$
and $p^I = (\tfrac{1}{2} + n_1, \ldots, \tfrac{1}{2} + n_8)$ with $n_i
\in \mZ$ and $\sum_i n_i \in 2\mZ$. This implies that $q^I_{1,2} =
N\beta^I_{1,2}$ is a lattice vector.

Consistency of the model requires the so called level
matching condition
\begin{equation}
  \label{eq:level}
  \Phi^2 = \beta_1^2 + \beta_2^2 \mod \frac{2}{N}
\end{equation}
to be satisfied \cite{OrbII,Vafa:1986wx}. It has long been suspected
\cite{PrivNilles, Aldazabal}\footnote{%
  In case of the $\Spin(32)/\mZ_2$ string theory such issues have been
  analyzed in \cite{Berkooz,Intriligator}. See also \cite{Aldazabal}.
  }, that this condition corresponds to the well known
relation\footnote{All traces for the gauge group will be in the
  adjoint representation.}
\begin{equation}
  \label{eq:GreenSchwarz}
  \tr R^2 = \frac{1}{30} \tr F_1^2 + \frac{1}{30} \tr F_2^2
\end{equation}
required by the Green-Schwarz mechanism \cite{GreenSchwarz} ($\Dd H =
\tr R^2 - 1/30 \tr F^2$) and the perturbativity of the orbifold ($H =
0$). However, for example by inspecting the different $\mZ_3$
orbifolds by trial and error, naive application leads to
inconsistencies.

Our main result, to be shown in section~\ref{Calc}, is that for an
$\pE_8$ bundle that corresponds to a shrunken instanton on a $\mZ_N$
orbifold singularity the fractional part of the instanton number is
given by\footnote{In analogy to the $\Spin{32}/\mZ_2$ case this
  relation has already been used in \cite{Aldazabal}. See the footnote
  in section \ref{INos} on how the $\pE_8$ instanton number is related
  to that of $\Spin(16)$.}
\begin{equation}
  \label{eq:Main}
  I = - \frac{1}{60} \frac{1}{8\pi^2} \int_U \tr F^2
  = \frac{N}{2} \beta^2
  \pmod 1
\end{equation}
where $\beta$ is defined as above. The rotation $r$ is given by
$\Phi=(1/N,-1/N)$ and $U$ is a small neighbourhood surrounding the
shrunken instanton. Using this result \eqref{eq:level} translates into
the requirement that locally the sum of the fractional parts of the
gauge instanton numbers match the fractional part of the gravitational
instanton number (which is $1/N$):
\begin{equation}
  \label{eq:level2}
  - \frac{1}{2} \frac{1}{8\pi^2} \int_U \tr R^2 =
  - \frac{1}{60} \frac{1}{8\pi^2} \int_U \tr F_1^2
  - \frac{1}{60} \frac{1}{8\pi^2} \int_U \tr F_2^2
  \pmod 1
\end{equation}
Even though this might look trivial from a weak coupling perspective,
in light of M-theory on $S^1/\mZ_2$ \eqref{eq:Main} fixes the
fractional instanton numbers on each $\mZ_2$ fixed point seperately.
Therefore the distribution of the integer part on the two $\mZ_2$
fixed points is not directly given by \eqref{eq:Main} and has to be
investigated by different methods \cite{Stieberger, OvrutI,
  Kaplunovsky, OvrutII}.

Turning to global aspects of the orbifold we define it in the usual
way \cite{OrbII} (for an introduction see \cite{Nilles} or \cite{Pol},
chapter 16) and explicitly allow for quantized Wilson lines.
Abstractly the orbifold is $O = \mC^2/S$ where $S$, the space group,
is generated by affine transformations $D : x \mapsto rx + h$.  To
include the gauge degrees of freedom we augment $D$ by a gauge
transformation $\gamma_D$ and require that the map $\gamma : S \auf
\pE_8 \times \pE_8$ is a group homomorphism. Furthermore we require
that all $\gamma_D$ commute with each other. This restricts the map
$\gamma$ quite severely and especially implies that each $\gamma_D$
generates a cyclic subgroup of $\pE_8$. Given a fixed point $Dx_0 =
x_0$ of some $D$ with $Dx=rx+h$ the twist is generated by $(r,
\gamma_D)$ which has to fulfill \eqref{eq:level}.

To see that $\gamma$ defines a flat gauge bundle consider
$\mC^2-F$ where $F$ is the set of all fixed points of $S$.  Since, by
supersymmetry, $F$ consists of isolated fixed points only the
fundamental group $\pi_1$ of $\mC^2-F$ is zero and $\mC^2-F$ is the
universal covering space of $(\mC^2-F)/S = O-F$, that is, the orbifold
with the fixed points taken out. This implies (see, for example
\cite{Bredon}, Chapter III) that $\pi_1$ of $O-F$ is isomorphic to $S$
and the map $\gamma$ provides a homomorphism of $\pi_1(O-F)$ to the
gauge group.  This is nothing but the data of a flat gauge
bundle\footnote{%
  Since on a flat bundle parallel translation around closed paths
  (with a fixed staring point) is invariant under continous
  deformations of the path, every class in $\pi_1$ corresponds to
  precisely one group element. Since concatenation of two pathes
  corresponds to group multiplication this is a homomorphism from
  $\pi_1$ to the gauge group.} on $O-F$ (up to gauge transformations).

In conclusion, we have shown that the orbifolds in our class
correspond to all possible flat $\pE_8 \times \pE_8$ bundles on the
orbifold $\mC^2/S$ with the fixed points taken out under the only
restriction that the sum of the fractional parts of the gauge
instanton numbers (computed from the flat bundle data via
\eqref{eq:level}) match the fractional part of the gravitational one
locally for every fixed point. Since the fractional instanton numbers
are computed seperately for every $\pE_8$, this classification fully
applies to M-theory on $S^1/\mZ_2$.

Moreover, \eqref{eq:Main} allows to calculate the fractional part of
an $\pE_8$ instanton number for any instanton sitting on a (possibly
blown up) orbifold singularity as long as the gauge bundle approaches
a flat bundle away from the singularity.

\section{Fractional Instanton Numbers}
\label{Calc}

This section is devoted to the derivation of equation \eqref{eq:Main}.
We consider the situation of an $\pE_8$ instanton localized in the
interior of a real four-manifold\footnote{%
  To be precise, we require $M$ to be orientable, compact and
  connected with a connected boundary.  } $M$ with boundary $L$, that
is we treat the gauge bundle as being flat on $L$. We start by showing
that the fractional part of the instanton number depends only on the
isomorphism class of the bundle on $L$. After that we construct one
explicit element of that class corresponding to the given data and
finally compute the fractional instanton number of that
element\footnote{%
  More mathematically speaking in step two we construct a bundle map
  from the Hopf fibration $S^3 \lauf S^2$ to $\pE_8 \lauf \pE_8/T^8$
  with $T^8$ a maximal torus of $\pE_8$ and in step three we calculate
  the image of the fundamental cycle of $S^3$ under that map by a
  spectral sequence.}.

The first two steps are mostly standard (see for example
\cite{Spanier, Steenrod}) whereas the third one requires a bit more
technology (see especially \cite{BottTu} chapter~III and \cite{Bott}).

\subsection{Basic Facts}

The instanton number is defined as (a discussion of the normalization
can be found in \cite{Atiyah:1978wi})
\begin{equation}
  \label{eq:INo}
  I = - \frac{1}{60} \frac{1}{8\pi^2} \int_M \tr F^2
\end{equation}
with the trace in the adjoint representation. Since $\pE_8$ is
semi-simple the first nontrivial homotopy group is $\pi_3(\pE_8) =
\mZ$.  This implies, due to the presence of the boundary, that the bundle
is trivial on $M$. This can be seen as follows: a possible obstruction
in constructing a section of a principal $\pE_8$-bundle on $M$ (and
thereby showing triviality of the bundle) is given by an nonzero
element of $H^4(M,\mZ)$. Since $M$ is an orientable and compact
manifold with boundary, we have by duality $H^4 (M,\mZ) \simeq H_0 (M,
\Dp M, \mZ) = 0$. (see for example \cite{Steenrod} part~III, or
\cite{Witten:1986bt} for a nice introduction) \eqref{eq:INo} now
reads
\begin{equation}
  \label{eq:INo2}
  I = - \frac{1}{30} \frac{1}{8\pi^2} \int_L
  \tr \left( AF - \frac{1}{3} A^3 \right)
  = \frac{1}{3\cdot 60} \frac{1}{8\pi^2} \int_L \tr A^3
\end{equation}

To make contact to the situation studied in the introduction we have
to take $L = S^3/\mZ_N$ (a lens space) with $\pi_1(L) = \mZ_N$. By
the same reasoning as in the introduction, $S^3$ is the covering space
of $L$ and $(r,\gamma)$ specifies a flat bundle on $L$.  Moreover, by
pulling back the bundle via the covering map $\pi : S^3 \auf
S^3/\mZ_N$ we get a bundle on $S^3$ on which $A$ can be gauge
transformed to $A'=0$ because $\pi_1(S^3)=0$. Denoting the gauge
transformation by $g: S^3 \auf \pE_8$ we get $A = g^{-1} A' g + g^{-1}
\Dd g = g^{-1} \Dd g$.  Plugging that into \eqref{eq:INo2} we get $I =
I_S/N$ with $I_S$ defined by
\begin{equation}
  \label{eq:INo3}
  I_S = \frac{1}{3\cdot 60} \frac{1}{8\pi^2}
  \int_{S^3} \tr (g^{-1}\Dd g)^3
\end{equation}
Of course $g$ represents an element of $\pi_3(\pE_8)$ which is
identified with $\mZ$ by \eqref{eq:INo3} (see again
\cite{Atiyah:1978wi}).  This especially shows that $I$ is a multiple
of $1/N$.

To show that $I_S$ changes by a multiple of $N$ when the bundle on $L$
is gauge transformed we consider two flat connections $A_1$ and $A_2$
on $L$, both corresponding to the same generator $(r,\gamma)$. Since
both bundles are isomorphic we have $A_2 = h^{-1} A_1 h + h^{-1} \Dd
h$ for some $h : L \auf \pE_8$. Now $h$ can be lifted to $S^3$ giving
$h_S = h \circ \pi$ and we get $A_2 = g_2^{-1} \Dd g_2 = h_S^{-1}
g_1^{-1} (\Dd g_1) h_S + h_S^{-1} \Dd h_S = (g_1 h_S)^{-1} \Dd (g_1
h_S)$. But since the pointwise product $g \cdot g'$ for two elements
of $\pi_3$ of a Lie group $G$ corresponds to the group addition of the
elements $g_2 = g_1 + h_S$.

To proceed, we note that, since $\pi_3$ is the first nontrivial
homotopy class of $\pE_8$, by the Hurewicz isomorphism $\pi_3(\pE_8)$
is isomorphic to $H_3(\pE_8, \mZ)$ and further by the universal
coefficient theorem to $H^3(\pE_8, \mZ)$ because, by the same argument
we have $H^1 = H^2 = H_1 = H_2 = 0$. Therefore the instanton number
$I_S$ is given by the pullback via $g$ of the generator $\omega$ of
$H^3(\pE_8, \mZ)$ evaluated on the fundamental cycle $C_S$ of $S^3$:
\begin{equation}
  \label{eq:Ino4}
  I_S = g^* \omega (C_S)
\end{equation}
Since $\pi : S^3 \auf L$ is $N$ to one we have $\pi_* C_S = N C_L$
with $C_L$ the fundamental cycle of $L$. This immediately yields $h_S^*
\omega (C_S) = \pi^* h^* \omega (C_S) = h^* \omega (\pi_* C_S)
= N h^* \omega (C_L) \in N \mZ$.

\subsection{Construction of the Bundle}

We now construct a bundle on $S^3$ simply by constructing the map $g$,
which we require to obey $g(rx) = \gamma g(x)$. This bundle will
obviously be the pullback of a bundle on $L$ that fulfills our
requirements.

To construct $g$ (and for the final step) we will need some basic
facts about lens spaces (see \cite{BottTu}, \S~18). As above we define
$L$ by the fibration $\mZ_N \lauf S^3 \overset{\pi}{\lauf} L$ where
$S^3$ is identified with the unit sphere in $\mC^2$ and the generator
of $\mZ_N$ acts on $S^3$ like
\begin{equation}
  \label{eq:ZN}
  e^{2\pi i/N} : (Z^1,Z^2) \mapsto (e^{2\pi i/N} Z^1, e^{2\pi i/N} Z^2)
\end{equation}
This action is of course compatible with the $\pU(1)$ action on $S^3$ (where
$\pU(1)$ is identified with the unit circle in $\mC$)
\begin{equation}
  \label{eq:U1}
  (Z^1,Z^2) \mapsto (\lambda Z^1, \lambda Z^2)
  \qquad
  \lambda \in S^1 \subset \mC
\end{equation}
and we get the Hopf-fibration $S^1 \lauf S^3 \overset{\pi_S}{\lauf}
\CP^1 \simeq S^2$. Now since $\mZ_N \subset \pU(1)$ the $\pU(1)$
action descends to an action on $L$ and we have
\begin{equation}
  \label{eq:BundleMap}
  \begin{matrix}
    S^1 & \lauf & S^3 & \overset{\pi_S}{\lauf} & S^2 \\
        &       & \hbox to 0pt {$\hss\pi$} \downarrow
        && \hbox to 0pt {$\hss\id$} \downarrow \\
    S^1 & \lauf & L & \overset{\pi_L}{\lauf} & S^2 \\
  \end{matrix}
\end{equation}
In this diagram $\pi$ is a bundle map which is $N$ to one on the
standard fibre.

This can be made more explicit by identifying $S^2$ with a cylinder $I
\times S^1$ where at both ends of the interval $S^1$ is identified to
a point. We parametrize $I \times S^1$ by $(\rho,\Psi)$ where $\rho
\in \left[ 0, \tfrac{\pi}{2} \right], \Psi \in \left[ 0, 2\pi
\right[$. To write down the bundle explicitly we divide $S^2$ into
upper ($D^2_+$) and lower ($D^2_-$) hemisphere. Using $(Z^1, Z^2) =
(\cos \rho \; e^{i\phi_1}, \sin \rho \; e^{i\phi_2})$ we write
\begin{equation}
  \label{eq:VS}
  \begin{aligned}
    D^2_+ &:& \rho &\geq \tfrac{\pi}{4}
    \qquad & \phi_2&=\phi_+ \qquad & \phi_1=\phi_+ - \Psi \\
    D^2_- &:& \rho &\leq \tfrac{\pi}{4}
    \qquad & \phi_1&=\phi_- \qquad & \phi_2=\phi_- + \Psi \\
  \end{aligned}
\end{equation}
Therefore $\lambda = e^{i\phi} \in \pU(1)$ acts like $\phi_- \mapsto
\phi_- + \phi, \; \phi_+ \mapsto \phi_+ + \phi$ and the fibre $S^1$ is
parametrized by $\phi_-, \phi_+ \in \left[ 0, 2\pi \right[$.

Since on the equator $\phi_+ = \phi_- + \Psi$ the sphere $S^3$ as a
bundle ist made of two trivial $S^1$ bundles on the hemispheres
clutched together by the generator of $\pi_1(S^1)$. Analogously, by
\eqref{eq:BundleMap}, the clutching function of $L$ is $N \in
\pi_1(S^1)$.

To simplify the construction of $g$ we insert a cylinder $C = I \times
S^1$ parametrized by $(x,\Psi)$, $x \in [0 , 1], \Psi \in \left[ 0,
  2\pi \right[$ between the hemispheres by attatching $D^2_+$ at $x=1$
and $D^2_-$ at $x=0$.  This is extended to the bundle by parametrizing
the fibre as $\phi_+$ as on $D^2_+$. Therefore the bundle is now
clutched nontrivially at $x=0$ and trivially at $x=1$.

Finally we define $g(\rho,\Psi,\phi_+) = g_+(\rho,\Psi,\phi_+) =
e^{iqH\phi_+}$ on $D^2_+$ and $g(\rho,\Psi,\phi_-) =
g_-(\rho,\Psi,\phi_-) = e^{iqH\phi_-}$ on $D^2_-$.  This yields
\begin{equation}
  \begin{array}{lll}
    g(x=0,\Psi,\phi_+)
    &= g_-(\rho=\tfrac{\pi}{4},\Psi,\phi_+ - \Psi)
    &= e^{iqH(\phi_+ - \Psi)} \\
    g(x=1,\Psi,\phi_+)
    &= g_+(\rho=\tfrac{\pi}{4},\Psi,\phi_+)
    &= e^{iqH\phi_+}
  \end{array}
\end{equation}
on the cylinder $C$ and $g$ can be extended to $g(x, \Psi, \phi_+) =
e^{iqH\phi_+}$ at $\Psi=0,2\pi$.

$g$ at $\phi_+=0$ is now defined on the boundary of the square $0\leq
x \leq 1, \; 0 \leq \Psi \leq 2\pi$ (the identification of $\Psi=0$
with $\Psi=2\pi$ plays no role in the consideration) and the only
nontrivial step in the construction is to extend this definition to
the interior of the square. Topologically this is the same as to
extend a map $g : S^1 \auf \pU(1) \subset \pE_8$ with $S^1 = \Dp D^2$
to the whole of $D^2$. Therefore we consider the fibration $\pU(1)
\lauf \pE_8 \lauf \pE_8/\pU(1)$ where $\pU(1)$ is the subgroup of
$\pE_8$ generated from $e^{iqH\phi}$. The relative homotopy sequence
of this fibration contains the following part
\begin{equation}
  \label{eq:RelHom}
  \ldots \lauf \pi_2(\pE_8) = 0 \lauf \pi_2 (\pE_8, \pU(1))
  \overset{\Dp}{\lauf}
  \pi_1(\pU(1)) \lauf \pi_1(\pE_8) = 0 \lauf \ldots
\end{equation}
Since the sequence is exact the map $\Dp$ is an isomorphism and $g$
can be extended to $\phi=0$ in $C$. Finally we define $g$ on the whole
of $C$ by $g(x,\Psi,\phi_+) = e^{iqH\phi_+} g(x,\Psi,\phi_+=0)$.

\subsection{Calculation of the Instanton Number}
\label{INos}

\newcommand{\dt}{\begin{rotate}{-30} $\xrightarrow{\ \ \ \ \ }$ \end{rotate}}

As the map $g$ constructed in the last section actually fulfills
$g(e^{i\phi}Z^i)= e^{iqH\phi} g(Z^i)$ it is a bundle map from $S^3$ to
$\pE_8$ which especially provides a homomorhism from the (cohomology)
spectral sequence of the former to that of the latter\footnote{It can
  be easily seen that we could restrict ourselves to a map to
  $\Spin(16)$ at this point: by acting with the Weyl group we can map
  $q$ into the $\Spin(16)$ sublattice of $\pE_8$. Since
  $\pi_1(\Spin(16))=0$ all steps of the last section apply as before
  and we are left with a pure $\Spin(16)$ bundle. This then allows to
  compute the instanton number with the formulas given in
  \cite{Intriligator, Aldazabal}. However, we will proceed in a
  different way, since our calculation gives the instanton number
  (including the integer part) for gauge bundles satisfying
  $g(e^{i\phi}Z^i)= e^{iqH\phi} g(Z^i)$.  }.

We start by writing down the standard example of the spectral sequence
of the Hopf fibration. The $E_2$ term is given by $E_2^{p,q} = H^p
(S^2) \otimes H^q (S^1)$, explicitly:
\begin{equation}
  \label{eq:E2S}
  E_2 = \;
  \begin{array}{c|ccc}
    1 & \mZ/a_S \dt & 0 & \mZ/a_Sb \\
    0 & \mZ/1       & 0 & \mZ/b \\
    \hline
      &  0          & 1 & 2 \\
  \end{array}
\end{equation}
where $p,q$ label columns and rows and the diagonal arrow denotes the
map $\Dd_2$. By $H/h$ we denote the group $H$ generated by $h$. As the
sequence stops at $E_3$ we have $E_3 \simeq H^*(S^3) = (\mZ/1, 0, 0,
\mZ/c_S)$ with $c_S = a_S b$. This implies that $\Dd_2$ is an
isomorphism from $E_2^{0,1}$ to $E_2^{2,0}$.

We now turn to the spectral sequence of the fibration $T^8 \lauf \pE_8
\lauf \pE_8/T^8$ where $T^8$ is a maximal torus of $\pE_8$ containing
the $\pU(1)$ generated from the elements $e^{iqH\phi}$. First we need
to verify that the base is simply connected. This is clear from the
homotopy sequence
\begin{equation}
  \label{eq:EHomotopy}
  \ldots \lauf \pi_1(\pE_8) =0 \lauf \pi_1(\pE_8/T^8) \lauf
  \pi_0(T^8) = \{*\} \lauf \cdots 
\end{equation}
($\pi_0$ is not a group here and consists only of the (arbitrary) base
point $*$). Furthermore, as shown by Morse theoretic methods in
\cite{Bott}, the base is torsion free and $H^{2n+1} (\pE_8/T^8) = 0$
for $n \in \mZ$. With this information the $E_2$ term is\footnote{%
  Addition and multiplication of groups are written with respect to the
  operations $\oplus$ and $\otimes$ on abelian groups, i.e.\ $2\mZ =
  \mZ \oplus \mZ$, $\mZ^2 = \mZ \otimes \mZ$.}
\begin{equation}
  \label{eq:E2}
  E_2 = \;
  \begin{array}{c|ccccc}
    3 & 56\mZ      &   &          &   & \\
    2 & 28\mZ \dt  & 0 &          &   & \\
    1 & 8\mZ  \dt  & 0 & 64\mZ\dt &   & \\
    0 & \mZ        & 0 & 8\mZ     & 0 & 35\mZ \\
    \hline
      &  0         & 1 & 2        & 3 & 4      \\
  \end{array}
\end{equation}
This can be seen as follows: Firstly, we note that $H^*(T^8) = (\mZ,
8\mZ, 28\mZ, 56\mZ, \ldots)$. Secondly, since $H^*(\pE_8) = (\mZ/1, 0, 0,
\mZ/\omega, \text{higher groups})$ by Hurewicz $E_3$ must be trivial
for degree 1 and 2. Therefore, again $\Dd_2 : E_2^{0,1} \auf
E_2^{2,0}$ is an isomorphism. This implies, since by the K\"unneth
formula $H^*(T^8) = (H^*(S^1)) ^8$ (with respect to $\otimes$), that
all maps $\Dd_2$ from the first to the third column are invertible.

Moreover, since $\Dd_3$ maps everything to zero all elements of
$H^*(\pE_8)$ up to degree three are given by $E_3$ up to degree three.
However, because $\Dd_2$ from the first to the third column is
invertible, the only nonvanishing element up to degree three of $E_3$
must be $E_3^{2,1} = \mZ/\omega$. This implies $E_2^{4,0} = 35\mZ$.

This can independently be verified by calculating $H^*(\pE_8/T^8)$ as
described in \cite{Bott}: the dimension of $H^{2n}(\pE_8/T^8)$ is
given by the number of elements of the Weyl group which change the
sign of precisely $q$ of the positive roots. This can be computed
easily, since, for a given Weyl reflection $\sigma$, $q$ is given by
the length of $\sigma$ (see for example \cite{Humphreys} section
10.3), i.e.\ the (minimal) number of simple Weyl reflections $\sigma$
can be composed of $\sigma = \sigma_{\alpha_1} \cdot \ldots \cdot
\sigma_{\alpha_q}$. For $q=1$ there are 8 simple roots so
$H^2(\pE_8/T^8) = 8\mZ$. For $q=2$ there are $8\cdot 7/2 + 7 = 35$
combinations, because simple Weyl reflections of two simple roots which
are not connected by a line in the Dynkin diagram commute. So
$H^4(\pE_8/T^8) = 35\mZ$.

Explicitly we denote the generators of $H^1(T^8)$ by $a^i, i=1,
\ldots, 8$ and those of $H^2(\pE_8/T^8)$ by $b^i = \Dd_2 a^i$. As
$E_2^{0,2} = H^2(T^8)$ is generated by the 28 elements $a^i a^j$ with
$i<j$ their image under $\Dd_2$ in $E_2^{2,1}$ is given by the
elements $\Dd_2 (a^i a^j) = b^i a^j - a^i b^j = - a^i b^j + a^j
b^i$.

To calculate $\omega$ we write all elements in terms of an euclidian
basis $A_I$ of the real homology $H_1(T^8,\mR)$ and its dual $a^I$ in
$H^1(T^8,\mR)$. We choose the basis such that it is compatible with
the lattice $\Lambda_8$ given in the introduction.  Then $\omega$ can
be written as $\omega = w_{IJ} a^I b^J$ where $w_{IJ}$ is a symmetric
matrix, because the image of $\Dd_2$ consists of all elements $w_{IJ}
a^I b^J$ with antisymmetric $w_{IJ}$.  Under an element $T$ of the Weyl
group $w$ is transformed to $T^{-1} w T$ ($T$ is an orthogonal
matrix). As $w$ must be invariant unter the Weyl group, which acts
irreducible on a vector, by Schur's lemma, $\omega= w \delta_{IJ} a^I
b^J$.

We now turn to the calculation of the instanton number.  By
construction $g$ maps the fundamental cycle $A_S$ of the $S^1$ fiber
of the Hopf fibration to $q^I A_I$. This implies
\begin{equation}
  g^* a^I (A_S) = a^I (g_* A_S) = a^I (q^J A_J) = q^I
\end{equation}
and therefore $g^* a^I = q^I a_S$. Finally we have
\begin{equation}
  \begin{split}
    I_S = g^* \omega (C_S)
    &= w \delta_{IJ} \; (g^* a^I \; g^* b^J) (C_S)
    = w \delta_{IJ} \; (g^* a^I \; g^* (\Dd_2 a^J)) (C_S) \\
    &= w \delta_{IJ} \; (g^* a^I \; \Dd_2 (g^* a^J)) (C_S)
    = w \delta_{IJ} \; (q^I a_S \; q^J \Dd_2 a_S) (C_S) \\
    &= w \delta_{IJ} \; q^I q^J \; (a_S b) (C_S)
    = w q^2
  \end{split}
\end{equation}
To normalize this equation we compare to the standard embedding
$I_S=1$, $q^I=(1,1,0,\ldots)$ resulting in $w=1/2$. Therefore we have
\begin{equation}
  \label{eq:Final}
  I = \frac{I_S}{N} = \frac{1}{2N} q^2 = \frac{N}{2} \beta^2 \pmod 1
\end{equation}

\section*{Acknowledgements}

This work was supported by the European Programs HPRN-CT-2000-00131,
HPRN-CT-2000-00148 and HPRN-CT-2000-00152.  I would like to thank
S.~F\"orste, G.~Honecker. H.~Klemm, R.~Schreyer, M.~Walter,
K.~Wendland, M.~Zucker and especially H.-P.~Nilles for many fruitful
discussions. I also thank A.~Font and A.~Uranga for bringing
\cite{Aldazabal} to my attention.

\end{document}